\documentclass[preprint,10pt,nofootinbib]{revtex4}
\usepackage[paper=letterpaper,margin=1.5in]{geometry}

\usepackage{epsfig}
\usepackage{graphics}\graphicspath{{graph/}{/eps}{}}
\usepackage{amssymb,amsmath}
\usepackage{time}
\newcommand{\p}{\partial}
\newcommand{\pa}{\partial}

\renewcommand{\vec}[1]{\textnormal{\boldmath$#1$}}
\newcommand{\vecc}[1]{\textnormal{\boldmath$#1$}}

\begin{document}
\begin{flushright}
{\normalsize
SLAC-PUB-12369\\
DESY-07-022\\
February 2007}
\end{flushright}

\title{Optical approximation in the theory of geometric
impedance\footnote{Work supported by Department of Energy contract
DE--AC02--76SF00515 and and by the EU contract 011935 EUROFEL.}}

\author{G. Stupakov, K.L.F. Bane}
\affiliation{Stanford Linear Accelerator Center, Stanford
University, Stanford, CA 94309}
\author{I. Zagorodnov}
\affiliation{Deutsches Elektronen-Synchrotron, Notkestrasse 85,
22603 Hamburg, Germany\vspace{.4cm} }
% \now, \today

\vspace{.4cm} \begin{abstract} In this paper we introduce an optical
approximation into the theory of impedance calculation, one valid in
the limit of high frequencies. This approximation neglects
diffraction effects in the radiation process, and is conceptually
equivalent to the approximation of geometric optics in
electromagnetic theory. Using this approximation, we derive
equations for the longitudinal impedance for arbitrary offsets, with
respect to a reference orbit, of source and test particles. With the
help of the Panofsky-Wenzel theorem we also obtain expressions for
the transverse impedance (also for arbitrary offsets). We further
simplify these expressions for the case of the small offsets that
are typical for practical applications. Our final expressions for
the impedance, in the general case, involve two dimensional
integrals over various cross-sections of the transition. We further
demonstrate, for several known axisymmetric examples, how our method
is applied to the calculation of impedances. Finally, we discuss the
accuracy of the optical approximation and its relation to the
diffraction regime in the theory of impedance. \vfill \centerline
{Submitted to Physical Review Special Topics--Accelerators and
Beams}
\end{abstract}
\maketitle

%*************** new section ********************************

\section{Introduction}

%*************** new section ********************************

The calculation of the impedance for the elements of a vacuum
chamber system and the associated calculation of beam dynamics
effects, such as beam instabilities or wakefield induced emittance
growth, are important elements in the design of a modern
accelerator. Sophisticated computer programs are routinely used for
such calculations, and in many cases they can successfully treat
complicated geometries, like those found in real vacuum systems.
There are, however, cases where the simulations are pushed to their
limits in resource requirements of both memory and cpu time, and in
their ability to yield an accurate result.

One example is the case of very short bunches, like those envisioned
in future linear colliders and future light sources. For example,
the final bunch length in the current design of the International
Linear Collider (ILC) \cite{yokoya06} is 300 microns; the final
(rms) bunch length in the Linac Coherent Light Source
\cite{lclsDesign02} is 20 microns and in the European XFEL
\cite{TeslaDesign02} it is 25 microns. Since numerical calculation
of the short-range wake requires a spatial mesh size equal to a
fraction of the bunch length, submillimeter bunches represent a
challenging computational task. Another example where direct
numerical calculation is difficult is related to long, small-angle
tapers which are often used to minimize the abruptness of vacuum
chamber transitions. For example, collimators with such tapers will
be used in the post-linac collimation section of the ILC. In such
cases numerical solution of Maxwell's equations requires a large
number of mesh points to fully cover the length of the transition or
collimator. The difficulty becomes especially pronounced for short
bunches.

The difficulty in both examples mentioned above is associated with a
small parameter. For short bunches such a small parameter is the
ratio of bunch length $\sigma_z$ to typical size $b$ of the
structure (in the vacuum chamber) that generates the impedance. If
we denote by $\lambdabar$ the inverse wavenumber $c/\omega$ ($c$ is
the speed of light and $\omega$ the characteristic frequency of
interest) then $\lambdabar \sim \sigma_z$, and the small parameter
for the problem is $\lambdabar/b$. It has long been known that
effective utilization of this small parameter may allow one to
simplify the impedance problem, and several analytical results are
available in the literature for the impedance at high-frequencies.
They include the impedance of a step transition
\cite{heifets91k,gianfelice90p} and the \emph{diffraction model} for
the impedance of a cylindrical pillbox cavity \cite{lawson90,
palmer90,bane90s,heifets90k}.
% when the length of cavity satisfies the
%relation $l_\mathrm{cav} \ll b^2/\lambdabar$ with $b$ the radius of
%the entrance pipe to the cavity.
More recently, a \emph{parabolic equation} method was developed that
provides a simplified treatment of diffraction effects at
high-frequencies \cite{stupakov06}.

In the case of diffraction theory, the calculation takes into
account the fact that radiated electromagnetic fields do not
propagate along straight lines. A Fresnel type integral from the
diffraction theory of light is used to evaluate the electromagnetic
energy that enters into the cavity region \cite{lawson90, palmer90}.
This energy is associated with the energy lost by the beam and is
thus related to the real part of the impedance. In this paper we
will show that in many cases the same kind of argument can be
applied to the calculation of impedance in an approximation that we
call the \emph{optical approximation}  (or \emph{optical regime}) in
the theory of impedance. In this approximation we assume that the
electromagnetic fields carried by a short bunch propagate along
straight lines equivalent to rays in the geometric optics. An
obstacle inside the beam pipe can intercept the rays and reflect
them away from their original direction. The energy in the reflected
rays is associated with the energy radiated by the beam, which can
then be related to the impedance. Note that this kind of argument
has been used in the past in the case of step-in and step-out
transitions in a round pipe, where the impedance was related to the
energy ``clipped away'' from the beam by the step
\cite{balakin83n,heifets90k}. In a recent paper this approach has
been extended and applied to the calculation of impedance for 3D
collimators \cite{zagorodnov06b}.

If $l$ is the length of an obstacle and $b$ is the minimal
cross-section size of the beam pipe, the conditions for the
optical approximation are
    \begin{align}\label{applicability_cond}
    \lambdabar
    \ll
    b
    \,,
    \qquad
    l
    \ll
    \frac{b^2}{\lambdabar}
    \,.
    \end{align}
The first of these two conditions requires the size of the obstacle
be much larger than the reduced wavelength of the radiation. The
right hand side of the second inequality has a meaning of the length
over which diffraction effects become significant, and this relation
guarantees that such effects give only a small correction to those
of geometric optics. Note that even a small-angle taper of angle
$\theta \sim b/l$ can be described in the optical approximation for
short enough bunches, if $\lambdabar/b \ll \theta$. The quantity
${b^2}/{\lambdabar}$ can also be interpreted as a catch-up distance
over which radiation, generated by the head of a beam, reflects from
a side wall of radius $b$ and reaches the beam tail at length
$\sigma_z \sim \lambdabar$ behind the head. Thus the second
condition of Eq.~(\ref{applicability_cond}) for the applicability of
the optical approximation is that the object is short compared to
the catch-up distance.

\begin{figure}[!htb]
    \centering
    \includegraphics[draft=false, width=0.4\textwidth]{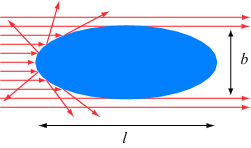}
    \caption{In geometric optics, light incident on a metallic object
    from the left is reflected by the surface creating a shadow behind the
    object. The reflected light can be considered as radiation emitted by the
    object due to currents induced in the metal by the incident electromagnetic field.}
    \label{fig:geometric_optics}
    \end{figure}

An analogous problem in geometric optics would be a body with
transverse size $b$ and length $l$ illuminated by light, see
Fig.~\ref{fig:geometric_optics}.
The energy reflected by the body can be simply calculated, in the
optical approximation, as the energy incident upon the
cross-sectional area of the body. Such calculation, however, is only
valid if the length of the obstacle $l$ and the transverse size $b$
obey the second condition in Eq.~(\ref{applicability_cond}). In the
opposite limit ($l\gg b^2/\lambdabar$), diffraction effects become
dominant.

In this paper we assume perfect conductivity in the vacuum chamber
wall. Our goal is to justify the optical approximation at
high-frequencies and to demonstrate how one can apply it to
calculations of both the longitudinal and transverse geometric
impedances. We take two approaches to the problem. The first one is
based on the general energy balance equation in electromagnetic
theory, which relates the longitudinal impedance to the energy
radiated from a transition in a beam pipe. In the past, calculations
of impedance based on this relation were carried out by one of the
authors in axisymmetric \cite{stupakov98_1} as well as rectangular
\cite{stupakov96_3} geometries. The derivation is similar to the
approach developed in Ref.~\cite{huang04sz} for the problem of laser
acceleration in vacuum. As it turns out, in the most general case of
unequal offsets of the leading and trailing particles, this approach
gives the sum of the longitudinal impedances symmetrized over the
coordinates of the particles, which does not provide enough
information to obtain the transverse impedance for the transition.
Our second approach uses a so-called indirect integration method
developed in Refs.~\cite{zagorodnov06,henke06b}. Although not as
physically transparent as the energy balance method, this
method---with some reasonable assumptions regarding the formation
length of the wake field---gives a simple expression for the
longitudinal impedance in the most general case.

We want to emphasize here that although our result is not derived
formally from first principles, it is based on a combination of
exact consequences of Maxwell's equations and simple physical
arguments that follow from the geometric optics. Our result is
applicable to an arbitrary three dimensional transition and allows
the incoming and outgoing beam pipes to have different
cross-sections. Practically important examples of impedance in the
optical regime are considered in a companion paper \cite{bane07sz}
where we also make a detailed comparison with computer simulations
and find excellent agreement between the theory and simulations.

This report is organized as follows. In Section \ref{sec:derivation}
we derive equations that relate the longitudinal impedance to the
integrals of the Fourier transformed Poynting vector over surfaces
located far from the transition. In Section \ref{sec:static_fields}
we calculate the contribution to the impedance of the static fields
carried by the charges while they are in the incoming and outgoing
pipes and far from the transition. The derivation in these two
sections is general and does not make any high-frequency
assumptions. In Section \ref{sec:radiation_field} the contribution
to the impedance of the radiation field is computed in the optical
approximation. In Section \ref{sec:indirect_integration} we invoke
the indirect integration method
to find the wake field for unequal offsets of the particles. In
Section \ref{sec:transverse_imp}, using the Panofsky-Wenzel theorem,
we derive expressions for the transverse impedance in the optical
approximation. Although our results for the transverse impedance are
applicable for arbitrarily large transverse offsets, we further
specialize them for the case of the small offsets usually assumed in
practical applications. Example impedance calculations are given in
Section \ref{sec:examples} where we compute both the longitudinal
and transverse impedances of a short, round collimator, and round
step-in and step-out transitions. We show that our results for these
cases agree with known expressions found in the literature. In
Section \ref{sec:pillbox_cavity} we discuss the relation between the
optical regime and diffraction theory, and establish the accuracy of
the optical approximation. Finally, in Section \ref{sec:summary} we
summarize the main results of the paper.

We use Gaussian system of units throughout this paper; to convert
impedances to MKS units, one multiplies them by the factor
$Z_0c/4\pi$, with $Z_0=377$~$\Omega$.

%*************** new section ********************************

\section{Longitudinal impedance and the energy radiated by charged particles}\label{sec:derivation}

%*************** new section ********************************

It is well known that the real part of the longitudinal impedance is
related to the energy loss of the beam \cite{chao93}. Since we
assume perfect conductivity in the metallic walls, the losses are
due to the beam-induced radiation that propagates away from the
transition. In this section we derive a formula that relates the
impedance to integrals of the field over remote surfaces far from
the transition, and which is valid for a transition of general shape
and incoming and outgoing beam pipes of arbitrary cross-section.

\begin{figure}[!h]
    \centering
    \includegraphics[width=0.4\textwidth]{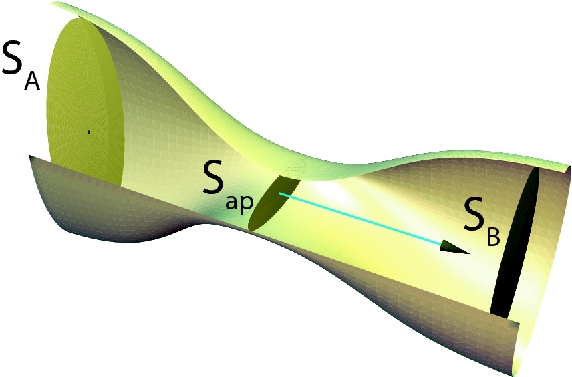}
    \caption{Geometry of a transition from pipe $A$ to pipe
    $B$. The source and test particles move from left to
    right along the $z$-axis. The aperture area
    $S_\mathrm{ap}$ is the minimal cross-section that connects the two pipes.
    \label{fig:transition}}
    \end{figure}

We consider transitions between two beam pipes of cross-sections
$S_A$ and $S_B$, respectively, as shown in
Fig.~\ref{fig:transition}. $S_A$ and $S_B$ are arbitrary, other
than their axes are parallel to each other and also parallel to
the so-called {\it design orbit} of the beam. We let the design
orbit, in turn, define the $z$-axis.
We assume that a charge $q_1$ moves at offset $\vec r_1 = (x_1,y_1)$
with respect to the design orbit, and a charge $q_2$ moves at offset
$\vec r_2= (x_2,y_2)$ and at distance $s$ behind (for $s>0$)  the
leading charge; both charges are assumed to move at the speed of
light $c$. We want to calculate the longitudinal wake
$w_\parallel(\vec r_1,\vec r_2,s)$ given by
    \begin{align}\label{wake}
    w_\parallel(\vec r_1,\vec r_2,s)
    =
    -\frac{c}{q_1}
    \int_{-\infty}^\infty E_{1,z}(\vec r_2,z=ct-s,t)dt
    \,,
    \end{align}
where $E_{1,z}(\vec r,z,t)$ is the $z$-component of the electric
field generated by the leading particle at the position of the
trailing one. The longitudinal impedance is related to the wake by
the Fourier transform
    \begin{align}\label{impedance}
    Z_\parallel(\vec r_1,\vec r_2,\omega)
    =
    \frac{1}{c}
    \int_{0}^\infty
    ds\,
    w_\parallel(\vec r_1,\vec r_2,s)
    e^{i\omega s/c}
    \,.
    \end{align}

Let us denote the electric and magnetic fields of the leading
charge by ${\vec E}_1(\vec r,z,t)$ and ${\vec H}_1(\vec r,z,t)$,
and the electric and magnetic fields of the trailing charge by
${\vec E}_2(\vec r,z,t)$ and ${\vec H}_2(\vec r,z,t)$. The field
$({\vec E}_1,{\vec H}_1)$ is due to the charge density of the
first particle, $\rho_1=q_1\delta(z-ct)\delta(\vec r - \vec r_1)$,
and the field $({\vec E}_2,{\vec H}_2)$ is generated by the charge
density of the second particle, $\rho_2 =
q_2\delta(z-s-ct)\delta(\vec r - \vec r_2)$. The total field is
the sum $\vec E = {\vec E}_1 + {\vec E}_2$, $\vec H = {\vec H}_1 +
{\vec H}_2$.

Using the energy balance equation in electrodynamics
\cite{jackson} we can calculate the energy lost by both charges by
integrating the Poynting vector over remote boundaries $S_A$ and
$S_B$ located, respectively, far to the left of the incoming pipe
and far to the right of the outgoing pipe
    \begin{align}\label{radiated_energy}
    W
    =
    \frac{c}{4\pi}
    \int_{-\infty}^{\infty}
    dt
    \int_{S_A+S_B}
    \vec n
    \cdot
    (
    \vec E\times\vec H
    )
    d S
    \,.
    \end{align}
Here $\vec n$ is the unit vector perpendicular to the surface area,
and we use the short-hand notation $\int_{S_A+S_B}$ for the sum of
the integrals over $S_A$ and $S_B$. In this equation the vector
$\vec n$ is oriented in the outward direction---it is parallel
(antiparallel) to the $z$ axis on $S_B$ ($S_A$). A positive value of
$W$ means that the charges lose energy. The contribution of such
integrals over the metallic surface of the pipes and the transition
vanishes because we assume that there are no losses in the wall. It
follows from the energy balance equation that the radiated energy
given by Eq.~(\ref{radiated_energy}) is equal to minus the energy
change of both particles
    \begin{align}\label{energy_balance}
    W
    =
    -{cq_1}
    \int_{-\infty}^\infty E_{z}(\vec r_1,z=ct,t)dt
    -{cq_2}
    \int_{-\infty}^\infty E_{z}(\vec r_2,z=ct-s,t)dt
    \,.
    \end{align}
The first term in this equation is the energy loss of the leading
particle, and the second term is the energy loss for the trailing
one. Note that strictly speaking the integration in
Eq.~(\ref{energy_balance}) should only be performed over the region
between the surfaces $S_A$ and $S_B$; however, assuming that the
surfaces are located sufficiently far from the transition (to where
the interaction of the charges vanishes), we can extend the limits
of integration to infinity.

The electric field at the location of the leading particle is equal
to $E_{1,z}(\vec r_1,z=ct,t) + E_{2,z}(\vec r_1,z=ct,t)$, and the
electric field at the location of the trailing particle is $
E_{1,z}(\vec r_2,z=ct-s,t) + E_{2,z}(\vec r_2,z=ct-s,t)$. We can
cast the formula for $W$ as follows:
    \begin{align}\label{energy_balance1}
    W
    =
    q_1^2U_1 + q_1q_2U_2 + q_2^2U_3
    \,,
    \end{align}
where
    \begin{align}
    U_1
    &=
    -\frac{c}{q_1}
    \int_{-\infty}^\infty E_{1,z}(\vec r_1,z=ct,t)dt
    \nonumber\\
    U_2
    &=
    -\frac{c}{q_1}
    \int_{-\infty}^\infty E_{1,z}(\vec r_2,z=ct-s,t)dt
    -\frac{c}{q_2}
    \int_{-\infty}^\infty E_{2,z}(\vec r_2,z=ct,t)dt
    \nonumber\\
    U_3
    &=
    -\frac{c}{q_2}
    \int_{-\infty}^\infty E_{2,z}(\vec r_2,z=ct-s,t)dt
    \,.
    \end{align}
Since we assume that the particles move at the speed of light, due
to causality the trailing particle does not affect the leading one.
This means that $E_{1,z}(\vec r_2,z=ct-s,t)$ is not equal to zero
only if $s>0$, and, similarly, $E_{2,z}(\vec r_2,z=ct,t)$ does not
vanish only for $s<0$. This observation allows us to rewrite $U_2$
as
    \begin{align}\label{second_term}
    U_2
    =
    h(s)w_\parallel(\vec r_1,\vec r_2,s)
    +
    h(-s)w_\parallel(\vec r_2,\vec r_1,-s)
    \,,
    \end{align}
where $h(s)$ is the unit step function.

The next step is to take the Fourier transform in time,
introducing the fields $\vecc{\mathcal E}$ and $\vecc{\mathcal
H}$,
    \begin{align}\label{fourier}
    \left\{
    \begin{array}
    {r}
    \vecc{\mathcal E}(\vec r,z,\omega)\\
    \vecc{\mathcal H}(\vec r,z,\omega)
    \end{array}
    \right\}
    =
    \frac{1}{2\pi}
    \int_{-\infty}^\infty
    dt\,
    e^{i\omega t}
    \left\{
    \begin{array}
    {r}
    \vec{ E}(\vec r,z,t)\\
    \vec{ H}(\vec r,z,t)
    \end{array}
    \right\}
    \,.
    \end{align}
Similarly, we also define the Fourier components $\vecc{\mathcal
E}_1(\vec r,z,\omega)$ and $\vecc{\mathcal H}_1(\vec r,z,\omega)$
for the electromagnetic field of the first charge. Because the
fields $\vec{E}$ and $\vec{H}$ are real, we have the relations
    \begin{align}\label{complex_conjugate}
    \vecc{\mathcal E}(\vec r,z,-\omega)
    =
    \vecc{\mathcal E}^*(\vec r,z,\omega)
    \,,
    \qquad
    \vecc{\mathcal H}(\vec r,z,-\omega)
    =
    \vecc{\mathcal H}^*(\vec r,z,\omega)
    \,,
    \end{align}
with equivalent relations for $ \vecc{\mathcal E}_1$ and $\vecc{
\mathcal H}_1$. The asterisk in these equations denote complex
conjugation. As for the Fourier components of the second charge, we
define them in a way that explicitly separates the phase factor
introduced by the distance $s$ between the particles:
    \begin{align}\label{fourier2}
    \left\{
    \begin{array}
    {r}
    \vecc{\mathcal E}_2(\vec r,z,\omega)\\
    \vecc{\mathcal H}_2(\vec r,z,\omega)
    \end{array}
    \right\}
    =
    \frac{1}{2\pi}
    e^{-i\omega s/c}
    \int_{-\infty}^\infty
    dt
    e^{i\omega t}
    \left\{
    \begin{array}
    {r}
    \vec{ E}_2(\vec r,z,t)\\
    \vec{ H}_2(\vec r,z,t)
    \end{array}
    \right\}
    \,.
    \end{align}
We then have $\vecc{\mathcal E} = \vecc{\mathcal E}_1 + e^{i\omega
s/c}\vecc{\mathcal E}_2$ and $\vecc{\mathcal H} = \vecc{\mathcal
H}_1 + e^{i\omega s/c}\vecc{\mathcal H}_2$.

From the linearity of Maxwell's equations it follows that the
electromagnetic field $(\vecc{\mathcal E}_1,\vecc{\mathcal H}_1)$
is generated by the charge density $\hat \rho_1$ equal to the
Fourier image of $\rho_1$
    \begin{align}\label{charge_1}
    \hat\rho_1
    =
    \frac{1}{2\pi}
    \int_{-\infty}^\infty
    dt\,
    \rho_1
    e^{i\omega t}
    =
    \frac{q_1}{2\pi c}
    e^{i\omega z/c}
    \delta(\vec r-\vec r_1)
    \,,
    \end{align}
and the field $(\vecc{\mathcal E}_2,\vecc{\mathcal H}_2)$ is
generated by the charge density
    \begin{align}\label{charge_2}
    \hat\rho_2
    =
    \frac{1}{2\pi}
    e^{-i\omega s/c}
    \int_{-\infty}^\infty
    dt\,
    \rho_2
    e^{i\omega t}
    =
    \frac{q_2}{2\pi c}
    e^{i\omega z/c}
    \delta(\vec r-\vec r_2)
    \,.
    \end{align}
Note that because of the extra phase factor $e^{-i\omega s/c}$ in
the definition (\ref{fourier2}) the Fourier components
$\hat\rho_1$ and $\hat\rho_2$ are now ``in phase''. Since
$\hat\rho_2$ has the property $\hat\rho_2(\omega) =
\hat\rho_2^*(-\omega)$, the fields $(\vecc{\mathcal
E}_2,\vecc{\mathcal H}_2)$ satisfy the same relations as Eqs.
(\ref{complex_conjugate}).

Using Parseval's theorem for Fourier transforms we can express $W$
in Eq.~(\ref{radiated_energy}) in terms of $\vecc{\mathcal
E}(\omega)$ and $\vecc{\mathcal H}(\omega)$:
    \begin{align}\label{eq_for_energy1}
    W
    &=
    \frac{c}{2}
    \int_{-\infty}^\infty d\omega
    \int_{S_A+S_B}
    [\vecc{\mathcal E}(\omega)
    \times
    \vecc{\mathcal H}^*(\omega)]
    \cdot \, \vec n\, dS
    \,,
    \end{align}
or, equivalently, in terms of the fields of the first and second
particles
    \begin{align}\label{eq_for_energy2}
    W
    &=
    \frac{c}{2}
    \int_{-\infty}^\infty d\omega\int_{S_A+S_B}
    [\vecc{\mathcal E}_1(\omega)
    \times
    \vecc{\mathcal H}_1^*(\omega)]
    \cdot \, \vec n\, dS
    \nonumber\\
    &+
    \frac{c}{2}
    \int_{-\infty}^\infty d\omega\int_{S_A+S_B}
    \left[
    e^{-i\omega s/c}
    \vecc{\mathcal E}_1(\omega)
    \times
    \vecc{\mathcal H}_2^*(\omega)
    +
    e^{i\omega s/c}
    \vecc{\mathcal E}_2(\omega)
    \times
    \vecc{\mathcal H}_1^*(\omega)
    \right]
    \cdot \, \vec n\, dS
    \nonumber\\
    &+
    \frac{c}{2}
    \int_{-\infty}^\infty d\omega\int_{S_A+S_B}
    [\vecc{\mathcal E}_2(\omega)
    \times
    \vecc{\mathcal H}_2^*(\omega)]
    \cdot \, \vec n\, dS
    \,.
    \end{align}

We will now prove that each of the three terms on the right hand
side of Eq.~(\ref{energy_balance1}) is equal to the corresponding
term in Eq.~(\ref{eq_for_energy2}). The proof is based on the
linearity of Maxwell's equations and the fact that expressions
(\ref{energy_balance1}) and (\ref{eq_for_energy2}) are equal for
arbitrary values of $q_1$ and $q_2$. If $q_2 = 0$, then there are
only the first terms in these two equations, hence they are equal.
If $q_1 = 0$, then there are only the third terms, and they are
also equal. Hence the second terms must be equal, too. Using
Eq.~(\ref{second_term}) we obtain
    \begin{align}\label{wake_in_terms_of_energy}
    &h(s)w_\parallel(\vec r_1,\vec r_2,s)
    +
    h(-s)w_\parallel(\vec r_2,\vec r_1,-s)
    =
    \nonumber\\
    &=
    \frac{c}{2q_1q_2}
    \int_{-\infty}^\infty d\omega\int_{S_A+S_B}
    \left[
    e^{-i\omega s/c}
    \vecc{\mathcal E}_1(\omega)
    \times
    \vecc{\mathcal H}_2^*(\omega)
    +
    e^{i\omega s/c}
    \vecc{\mathcal E}_2(\omega)
    \times
    \vecc{\mathcal H}_1^*(\omega)
    \right]
    \cdot \, \vec n\, dS
    \nonumber\\
    &=
    \frac{c}{2q_1q_2}
    \int_{-\infty}^\infty d\omega\,
    e^{-i\omega s/c}
    \int_{S_A+S_B}
    \left[
    \vecc{\mathcal E}_1(\omega)
    \times
    \vecc{\mathcal H}_2^*(\omega)
    +
    \vecc{\mathcal E}_2^*(\omega)
    \times
    \vecc{\mathcal H}_1(\omega)
    \right]
    \cdot \, \vec n\, dS
    \,.
    \end{align}
In the last integral we changed the integration variable
$\omega\rightarrow -\omega$ in the second term of the integrand
and used the relations $\vecc{\mathcal E}_2(-\omega) =
\vecc{\mathcal E}_2^*(\omega)$ and $    \vecc{\mathcal
H}_1^*(-\omega) = \vecc{\mathcal H}_1(\omega)$. Taking the Fourier
transform of this equation and using Eq.~(\ref{impedance}) gives
    \begin{align}\label{impedance_real1}
    {\cal Z}(\vec r_1,\vec r_2,\omega)
    &\equiv
    Z_\parallel(\vec r_1,\vec r_2,\omega)
    +
    Z_\parallel(\vec r_2,\vec r_1,\omega)
    \nonumber\\
    &=
    \frac{\pi c}{q_1q_2}
    \int_{S_A+S_B}
    \left[
    \vecc{\mathcal E}_1(\omega)
    \times
    \vecc{\mathcal H}_2^*(\omega)
    +
    \vecc{\mathcal E}_2^*(\omega)
    \times
    \vecc{\mathcal H}_1(\omega)
    \right]
    \cdot \, \vec n\, dS
    \,.
    \end{align}
The quantity ${\cal Z}$ in this equation is the sum of the
impedances symmetrized with respect to the offsets of the leading
and trailing particles. If $\vec r_1 = \vec r_2$, then the left hand
side is equal to $2Z_\parallel(\vec r_1,\vec r_1,\omega)$ and this
equation gives the longitudinal impedance for particles that have
the same offset. However, for the general case of unequal offsets,
the equation gives only the sum of the impedances ${\cal Z}$.

Note that up to this point we did not use any approximation in our
derivation---our result is valid not only in the optical regime, but
for arbitrary frequency $\omega$. Eq.~(\ref{impedance_real1}),
relates the sum of two longitudinal impedances to the interference
term between the electromagnetic fields of charges 1 and 2 in the
energy flow through the remote boundaries $S_A$ and $S_B$. In the
next sections we will split the right hand side of
Eq.~(\ref{impedance_real1}) into several distinct parts and then
calculate the parts individually in the optical regime.

%*************** new section ********************************

\section{Contribution to impedance of static and radiation
fields}\label{sec:static_fields}

%*************** new section ********************************

There are several contributions to the integral in
Eq.~(\ref{impedance_real1}). The first one comes from the field that
is carried by particles when they cross the surface $S_A$ on their
way to the transition, and the second one arises when they pass
through the surface $S_B$ after leaving the transition. We call
these fields the \emph{static} fields because they do not vary in
time in the beam frame and they can be calculated from a time
independent system of equations after a trivial change of variables
$z-ct\rightarrow \zeta$ in the laboratory frame. We emphasize here
that these are not radiation fields; nevertheless, they need to be
included into the original energy balance in
Eq.~(\ref{radiated_energy}).

We begin calculation of the static field contributions from pipe
$A$. The electric fields $\vecc{\mathcal E}_{1,A}$ and
$\vecc{\mathcal E}_{2,A}$ in the straight pipe can be represented
in terms of the potential function $\phi$,
    \begin{align}\label{static_E}
    \vecc{\mathcal E}_{1,A}
    =
    -
    \frac{q_1}{2\pi c}
    e^{i\omega z/c}
    \nabla \phi_{1,A}(\vec r)
    \,,
    \qquad
    \vecc{\mathcal E}_{2,A}
    =
    -
    \frac{q_2}{2\pi c}
    e^{i\omega z/c}
    \nabla \phi_{2,A}(\vec r)
    \,,
    \end{align}
where $\nabla = \vec {\hat x}\, \p/\p x + \vec {\hat y}\, \p/\p
y$. The potential $\phi$ satisfies Poisson's equation with source
terms given by the charge densities of Eqs. (\ref{charge_1}) and
(\ref{charge_2}):
    \begin{align}\label{eqs_for_phi}
    \nabla^2\phi_{1,A}(\vec r)
    =
    -
    4\pi\delta(\vec r-\vec r_1)
    \,,
    \qquad
    \nabla^2\phi_{2,A}(\vec r)
    =
    -
    4\pi\delta(\vec r-\vec r_2)
    \,,
    \end{align}
with boundary conditions $\phi_{1,A} = \phi_{2,A} =0$ on the wall
of pipe $A$. The magnetic field in pipe $A$ is: $\vecc{\mathcal
H}_{1,A} = \vec{\hat z}\times \vecc{\mathcal E}_{1,A}$ and
$\vecc{\mathcal H}_{2,A} = \vec{\hat z}\times \vecc{\mathcal
E}_{2,A}$. We denote by $Z_A$ the contribution of these fields,
when integrated over the surface $S_A$, to the integral
Eq.~(\ref{impedance_real1})
    \begin{align}\label{Z_a}
    Z_A
    &=
    -
    \frac{2\pi c}{q_1q_2}
    \int_{S_A}
    \vecc{\mathcal E}_1(\omega)
    \cdot
    \vecc{\mathcal E}_2^*(\omega)
    d S
    =
    -
    \frac{1}{2\pi c}
    \int_{S_A}
    \nabla \phi_{1,A}
    \cdot
    \nabla \phi_{2,A}
    \, d S
    \,.
    \end{align}
The minus sign in this equation is due to the fact that the normal
vector to $S_A$ is oriented in the direction opposite to the $z$
axis.

In the same way we calculate the contribution $Z_B$ to
Eq.~(\ref{impedance_real1}) from the remote boundary $S_B$. The
electric field on this surface $\vecc{\mathcal E}_{1,B}$ and
$\vecc{\mathcal E}_{2,B}$ is given by the same equations
(\ref{static_E}) and (\ref{eqs_for_phi}) with the index $A$
replaced by $B$, and the boundary conditions for $\phi_{1,B}$ and
$\phi_{2,B}$ being zero on the metallic surface of beam pipe $B$.
However, because the direction of the normal vector to $S_B$ is
along the $z$ axis, $Z_B$ has a positive sign
    \begin{align}\label{Z_b}
    Z_B
    =
    \frac{1}{2\pi c}
    \int_{S_B}
    \nabla \phi_{1,B}
    \cdot
    \nabla \phi_{2,B}
    \, d S
    \,.
    \end{align}
Note that, strictly speaking, the integrals in Eqs. (\ref{Z_a})
and (\ref{Z_b}) diverge because the field has a singularity at the
location of charges $q_1$ and $q_2$. However in the sum $Z_A+Z_B$
the singular contributions cancel, and the result is finite.

In addition to the terms $Z_A$ and $Z_B$ there will be a
contribution to the impedance due to the radiation field emitted
in the transition region. We will denote this contribution
$Z_\mathrm{rad}$, so that
    \begin{align}
    {\cal Z}
    =
    Z_A+Z_B
    +
    Z_\mathrm{rad}
    \,.
    \end{align}
The quantity $Z_\mathrm{rad}$ will be evaluated in the next section
using the optical approximation.

%*************** new section ********************************

\section{Radiation from the transition in the optical
approximation}\label{sec:radiation_field}

%*************** new section ********************************

In the optical regime the radiated energy has several terms. The
first term is the energy that is radiated by reflection from the
narrowing part of the pipe in the aperture area\footnote{For a 3D
transition with a complicated geometry the general rule for
finding the aperture area $S_\mathrm{ap}$ is the following. Assume
that the transition is illuminated by a set of parallel rays of
light that propagate from pipe A along the $z$ axis. Then the
cross-section of the illuminated area in pipe B gives
$S_\mathrm{ap}$.}\label{S_ap} $S_\mathrm{ap}$ (see
Fig.~\ref{fig:transition}). We denote the cross-section of the
transition that complements $S_\mathrm{ap}$ to $S_A$ by
$S_A-S_\mathrm{ap}$, and, similarly, the cross-section that
complements $S_\mathrm{ap}$ to $S_B$ by $S_B-S_\mathrm{ap}$. The
incident energy flux within this area will be ``clipped away''
from the beam and converted into a radiation field. This radiation
may go in the backward direction if the narrowing region is steep
enough (\textit{e.g.}, a diaphragm, or a 90-degree, abrupt step-in
pipe radius), or, for a small-angle taper, it may go in the
forward direction. Since the incident energy is the static field
in pipe $A$, the contribution $Z_\mathrm{rad,1}$ of the clipped
energy is given by the same Eq.~(\ref{Z_a}), but with the
integration now carried out over the ``clipping'' area
$S_A-S_\mathrm{ap}$,
    \begin{align}\label{Z_rad1}
    Z_\mathrm{rad,1}
    =
    \frac{1}{2\pi c}
    \int_{S_A-S_\mathrm{ap}}
    \nabla \phi_{1,A}
    \cdot
    \nabla \phi_{2,A}
    \, d S
    \,.
    \end{align}
The sign in this equation is positive because the radiation
propagates from the transition region to infinity.

Through the aperture $S_\mathrm{ap}$ connecting pipes $A$ and $B$,
charges carrying the static potential $\phi_{A}$  enter into pipe
$B$ (for brevity, we momentarily drop indices 1 and 2). The fields
will eventually change in such a way that at a large distance from
the transition the particles will carry the potential $\phi_{B}$.
In the course of this restructuring, there will be additional
radiation emitted in the forward direction. We denote the
contribution of this radiation to the impedance by
$Z_\mathrm{rad,2}$ and calculate it in the following way. We
represent the potential of the charges immediately after passing
through the aperture as a sum of the potential $\phi_B$ occupying
area $S_B$, the potential $-\phi_B$ over the area $S_B -
S_\mathrm{ap}$, and the potential $\phi_A-\phi_B$ over the area
$S_\mathrm{ap}$. From the linearity of Maxwell's equations, the
first field proceeds with the charges as a new static field in
pipe B, and the last two transform into radiation. The energy
integral of the interference fields carrying this radiation is
similar to Eqs. (\ref{Z_a}) and (\ref{Z_b}) with the difference
that over the area $S_\mathrm{ap}$ one needs to use
$\phi_A-\phi_B$, and outside of it, the potential $-\phi_B$:
    \begin{align}\label{Z_rad2}
    Z_\mathrm{rad,2}
    =
    \frac{1}{2\pi c}
    \int_{S_\mathrm{ap}}
    \nabla (\phi_{1,A}-\phi_{1,B})
    \cdot
    \nabla (\phi_{2,A}-\phi_{2,B})
    \, d S
    +
    \frac{1}{2\pi c}
    \int_{S_B - S_\mathrm{ap}}
    \nabla \phi_{1,B}
    \cdot
    \nabla \phi_{2,B}
    \, d S
    \,.
    \end{align}

\begin{figure}[!h]
    \centering
    \includegraphics[width=0.4\textwidth]{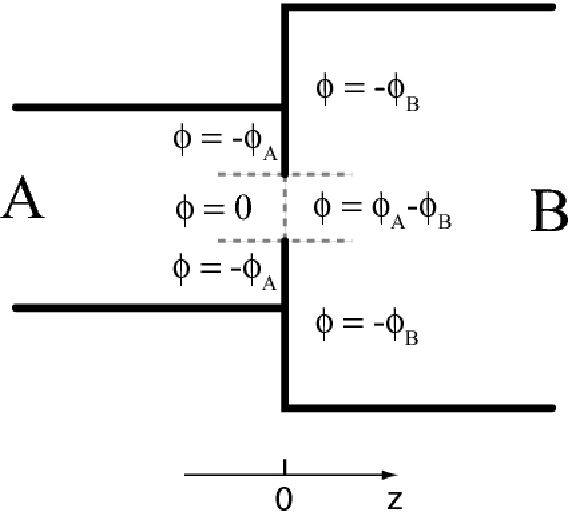}
    \caption{The transition of Fig. \ref{fig:transition}, represented schematically as an
    infinitely
    thin protrusion. Also given are the boundary conditions for
    the potentials connected to the radiation fields on both the A and B
    sides of the protrusion.
    \label{fig:transition1}}
    \end{figure}

A slightly different view of the derivation of Eqs. (\ref{Z_rad1})
and (\ref{Z_rad2}) is illustrated by Fig.~\ref{fig:transition1}.
According to this view, since the length of the transition is short
[to satisfy Eqs.~(\ref{applicability_cond})], it can be treated as
an infinitely short protrusion located at $z=0$. The incident
electromagnetic field in pipe A, given by potential $\phi_A$,
generates a radiation field which will be reflected back into pipe
A. The radiation field satisfies the boundary conditions $\phi =
-\phi_A$ on the left side of the protrusion, at $z=-0$, indicated in
Fig. \ref{fig:transition1}. This condition follows from the
requirement that the tangential electric field on the surface of the
metal is zero. The boundary conditions on the side of pipe B, at
$z=+0$, are also indicated in the figure. The radiation field that
satisfies these boundary conditions, when summed with the static
field corresponding to potential $\phi_B$, gives a field equal to
zero on the metallic surface of the protrusion, and equal to
$\phi_A$ over the aperture $S_{ap}$. The interference terms between
the radiation fields of the first and the second charges lead to the
integrals Eq.~(\ref{Z_rad1}) and (\ref{Z_rad2}) on the left and the
right side of the transitions, respectively.

Collecting now all contributions in Eqs. (\ref{Z_a}), (\ref{Z_b}),
(\ref{Z_rad1}) and (\ref{Z_rad2}) we arrive at the final result
    \begin{align}\label{Z_total}
    {\cal Z}
    =
    \frac{1}{\pi c}
    \int_{S_B}
    \nabla \phi_{1,B}
    \cdot
    \nabla \phi_{2,B}
    \, d S
    -
    \frac{1}{2\pi c}
    \int_{S_\mathrm{ap}}
    \left(
    \nabla \phi_{1,A}
    \cdot
    \nabla \phi_{2,B}
    +
    \nabla \phi_{1,B}
    \cdot
    \nabla \phi_{2,A}
    \right)
    \, d S
    \,,
    \end{align}
where the first integral is carried out over the cross-section of
pipe B, and the second integral over aperture $S_\mathrm{ap}$.
Note that the impedance given by Eq.~(\ref{Z_total}) is real and
does not depend on frequency $\omega$.

As was emphasized above, the energy balance equation which was used
to derive Eq. (\ref{Z_total}) gives the summed impedance ${\cal
Z}_\parallel(\vec r_1,\vec r_2,\omega) =Z_\parallel(\vec r_2,\vec
r_1,\omega) + Z_\parallel(\vec r_1,\vec r_2,\omega)$, which is
symmetrized over the offsets of the leading and trailing particles.
This suffices to give us the longitudinal impedance for the case
when leading and trailing particles follow the same path. However,
for the transverse impedance (which we will discuss below) we will
need to first find $Z_\parallel(\vec r_1,\vec r_2,\omega)$ alone
($\cal Z_\parallel$ is not sufficient). To find $Z_\parallel(\vec
r_1,\vec r_2,\omega)$ in the optical approximation, we will, in the
next section, invoke a more formal approach based on a so-called
indirect integration method.

%*************** new section ********************************

\section{Indirect integration method }\label{sec:indirect_integration}

%*************** new section ********************************

The indirect integration method developed in Refs.
\cite{zagorodnov06,henke06b} reduces the calculation of the wake
field to the integration along a straight path up to some point
$z_0$ located in the exit pipe B. The contribution from the
remaining path $z>z_0$ is expressed through the solution of an
auxiliary problem at the cross section $z=z_0$.

By choosing $z_0$ in pipe B immediately after the transition, we
note that, in the optical regime, the contribution of path $z<z_0$
can be neglected. Indeed, the wake accumulates over the catch up
distance $\sim b^2/\lambdabar$, which, according to Eqs.
(\ref{applicability_cond}), is much larger than the transition
length $l$. This observation greatly simplifies the problem and
allows us to compute the impedance from the part of path with
$z>z_0$ only. This contribution is \cite{zagorodnov06}
    \begin{align}\label{eq2}
    w(\vec{r}_1,\vec{r}_2,s)
    &=
    -
    \frac{1}{q_1}\Phi(\vec{r}_2,s)
    \,,
    \end{align}
where $\Phi$ satisfies the equation
    \begin{align}\label{eq2a}
    \nabla^2 \Phi(\vec{r}, s)
    &=
    \left(
    \frac{\p }{\p z}
    -
    c^{-1}
    \frac{\p}{\p t}
    \right)
    E_{1,z}^\mathrm{sc}(\vec{r},z=0,t=s/c)\ ,
    \end{align}
with the boundary condition $\Phi = 0$ on the wall of pipe B. The
quantity $E_{1}^\mathrm{sc}$ is (according to the terminology of
Ref. \cite{zagorodnov06}) the ``scattered'' electric field of the
leading charge---it is obtained by subtracting  from the total
field of this charge its static field in pipe B. In Eq.
(\ref{eq2a}) we set $z_0 = 0$ and suppress the argument $\vec r_1$
in the function $\Phi$---the dependence on this argument is clear
from Eq. (\ref{eq2a}) where the electric field $E_{1,z}$ is
generated by the leading particle moving with offset $\vec r_1$.

We now transform Eq. (\ref{eq2a}) in a way that eliminates the
longitudinal component $E_z$. Using $\p E_z^\mathrm{sc}/\p z +
\nabla\cdot \vec E_\perp^\mathrm{sc} = 0$ and $c^{-1}\p
E_z^\mathrm{sc}/\p t = (\nabla\times \vec{H}^\mathrm{sc})_z$ we
obtain
    \begin{align}\label{eq3}
    \frac{\p E_{1,z}^\mathrm{sc}}{\p z}
    &=
    -\nabla \cdot\vec{E}_{1\perp}^\mathrm{sc}
    \nonumber\\
    c^{-1}\frac{\p E_{1,z}^\mathrm{sc}}{\p t}
    &=
    (\nabla\times \vec{H}_{1\perp}^\mathrm{sc})_z
    \,,
    \end{align}
where the symbol $\perp$ refers to the components of the field
perpendicular to the $z$ axis (we also remind the reader that
$\nabla$ is a two-dimensional operator in the $x-y$ plane).

Because we now use the time domain representation for the fields, we
need to take the inverse Fourier transform of Eq. (\ref{static_E})
(and of a similar equation for pipe B) to find the static fields of
the particle. This gives for $\vec E_{1}$
    \begin{align}
    \vec{E}_{1,A}(\vec r,z,t)
    =
    -
    \frac{q_1}{c}
    \delta(t- z/c)
    \nabla \phi_{1,A}(\vec r)
    \,,
    \qquad
    \vec{E}_{1,B}(\vec r,z,t)
    =
    -
    \frac{q_1}{c}
    \delta(t- z/c)
    \nabla \phi_{1,B}(\vec r)
    \,,
    \end{align}
with the magnetic fields given by  $\vecc{H}_{1,A} = \vec{\hat
z}\times \vecc{ E}_{1,A}$ and $\vecc{H}_{1,B} = \vec{\hat z}\times
\vecc{ E}_{1,B}$.

The crucial step in the derivation is to notice that, in the
optical regime, after passage through the transition region, the
static field of particle 1 is ``scraped off'' outside the aperture
$S_\mathrm{ap}$. The field left with the charge is equal to
$\vecc{ E}_{1,A}$ but only within the area of $S_\mathrm{ap}$ (the
field in $S_B-S_\mathrm{ap}$ is zero). To find the scattered field
we need to subtract from the ``truncated'' field $\vec{E}_{1,A}$
the static field $\vec{E}_{1,B}$ of charge 1 in pipe B, which
gives
    \begin{equation}\label{eq4}
    \vec E_{1\perp}^\mathrm{sc}(\vec r,z,t)
    =
    \begin{cases}
    -
    ({q_1}/{c})
    \nabla
    [\phi_{1,A}(\vec r)-\phi_{1,B}(\vec r)]
    \delta(t-z/c)
    \,\,
    &
    \text{in }S_\mathrm{ap}
    \,,
    \\
    -
    ({q_1}/{c})
    \nabla
    \phi_{1,B}(\vec r)
    \delta(t-z/c)
    &
    \text{in }S_B-S_\mathrm{ap}
    \,,
    \end{cases}
    \end{equation}
with the corresponding magnetic field given by
    \begin{align}\label{eq5}
    \vec H_{1\perp}^\mathrm{sc}
    =
    \vec {\hat z} \times \vec E_{1\perp}^\mathrm{sc}
    \,.
    \end{align}
We substitute Eqs. (\ref{eq4}) and (\ref{eq5}) into Eq. (\ref{eq3})
and use the result as the right hand side in Eq. (\ref{eq2a}) for
$\Phi$. It follows from Eq. (\ref{eq5}) that $(\nabla\times \vec
H_{1\perp}^\mathrm{sc})_z = \nabla\cdot \vec
E_{1\perp}^\mathrm{sc}$, which gives
    \begin{align}
    \nabla^2 \Phi(\vec{r},s)
    =
    -2
    \nabla \cdot\vec E_{1\perp}^\mathrm{sc}(\vec{r},z=0,t=s/c)
    \,.
    \end{align}
This equation is solved by first noting that
$-(4\pi)^{-1}\phi_{B}(\vec r)$ is the Green function in $S_B$
($\phi_{B}$ satisfies Eqs. (\ref{eqs_for_phi}) in which A is
replaced by B) and then using the symmetry of the Green function
with respect to its two arguments \cite{morse_feshbach}. The result
is
    \begin{align}
    \Phi(\vec{r}_2,s)
    &=
    \frac{1}{2\pi}
    \int_{S_B}dS
    \phi_{2,B}(\vec r)
    \nabla \cdot\vec{E}_{1\perp}^\mathrm{sc}(\vec{r},z=0,t=s/c)
    \nonumber\\
    &=
    -
    \frac{1}{2\pi}
    \int_{S_B}dS
    \nabla\phi_{2,B}(\vec r)\cdot
    \vec{E}_{1\perp}^\mathrm{sc}(\vec{r},z=0,t=s/c)
    \nonumber\\
    &=
    -
    \frac{q_1}{2\pi}
    \delta(s)
    \left[
    -
    \int_{S_{ap}}
    dS\,
    \nabla\phi_{2,B}\cdot
    \nabla
    (\phi_{1,A}-\phi_{1,B})
    +
    \int_{S_B - S_{ap}}
    dS\,
    \nabla\phi_{2,B}\cdot
    \nabla\phi_{1,B}
    \right]
    \,,
    \end{align}
where in the second integral we integrated by parts and used the
fact that $\phi_{2,B}$ vanishes on the wall of pipe B. Our final
result for the wake becomes
    \begin{align}
    w
    =
    \frac{1}{2\pi}
    \delta(s)
    I
    \,,
    \end{align}
where
    \begin{align}\label{eq1a}
    I
    &=
    \int_{S_B}
    \nabla \phi_{1,B}(\vec r)
    \cdot
    \nabla \phi_{2,B}(\vec r)
    \,dS
    -
    \int_{S_{ap}}
    \nabla \phi_{1,A}(\vec r)
    \cdot
    \nabla \phi_{2,B}(\vec r)
    \,dS
    \,.
    \end{align}
The impedance corresponding to this wake is
    \begin{align}\label{eq1}
    Z(\vec{r}_1,\vec{r}_2)
    =
    \frac{1}{2\pi c}
    I
    \,.
    \end{align}
It is easy to see that this impedance is consistent with the
symmetrized formula Eq. (\ref{Z_total}).

%*************** new section ********************************

\section{Transverse impedance and small offset of
particles}\label{sec:transverse_imp}

%*************** new section ********************************

Knowledge of the longitudinal impedance allows one to compute the
transverse impedance using the Panofsky-Wenzel theorem
\cite{panofsky56w}. In the general case, the transverse impedance
is represented by a vector $\vec Z_\perp$ perpendicular to the
particle's orbit, and is given by
    \begin{align}\label{panofsky_wenzel}
    \vec Z_\perp
    =
    \frac{c}{\omega}
    \nabla_{\vec r_2} Z_\parallel
    \,,
    \end{align}
where $\nabla_{\vec r_2}$ is the operator ``nabla'' that
differentiates with respect to the coordinates $\vec r_2$ of the
trailing particle, $\nabla_{\vec r_2} = \hat {\vec x}\, \pa/\pa
x_2 + \hat {\vec y}\, \pa/\pa y_2$, with $\hat {\vec x}$ and $\hat
{\vec y}$ being the unit vectors in $x$ and $y$
directions\footnote{In this paper we use the following definitions
of the transverse wake $\vec w_\perp$
    \begin{align*}
    w_\perp(\vec r_1,\vec r_2,s)
    =
    \frac{c}{q_1}
    \int_{-\infty}^\infty
    \left[
    \vec E_{1,\perp}(\vec r_2,z=ct-s,t)
    +
    \hat{\vec z}
    \times
    \vec H_{1}(\vec r_2,z=ct-s,t)
    \right]
    dt
    \,,
    \end{align*}
and the transverse impedance
%    \begin{align*}
$
    \vec Z_\perp(\vec r_1,\vec r_2,\omega)
    =
    -(i/c)
    \int_{0}^\infty
    ds\,
    \vec w_\perp(\vec r_1,\vec r_2,s)
    e^{i\omega s/c}
    \,.
    $
%    \end{align*}
}.

In applications, it is typically assumed that there is a symmetry
axis in the system and the beam has a small offset relative to
this axis compared to the transverse size of the pipe. In this
case, one can expand the impedance in Taylor series. The leading
terms in the transverse impedance in this case are linear in
offsets of the leading and trailing particles\footnote{As is well
known, for axisymmetric systems the transverse impedance does not
depend on the offset of the trailing particle. This however is not
true for systems which are not axisymmetric.}. We will now derive
expressions for the transverse impedance in this approximation.

For small offsets of the leading and trailing particles we can
expand the delta functions in Eqs. (\ref{eqs_for_phi})
    \begin{align}
    \delta(\vec r-\vec r_1)
    &\approx
    \delta(\vec r)
    -
    \vec r_1\cdot\nabla\delta(\vec r)
    +
    \frac{1}{2}
    (\vec r_1\cdot\nabla)^2\delta(\vec r)
    \,,
    \nonumber \\
    \delta(\vec r-\vec r_2)
    &\approx
    \delta(\vec r)
    -
    \vec r_2\cdot\nabla\delta(\vec r)
    +
    \frac{1}{2}
    (\vec r_2\cdot\nabla)^2\delta(\vec r)
    \,,
    \end{align}
and represent each potential as a sum of a monopole part
$\phi^{(\mathrm{m})}$, a dipole part  $\phi^{(\mathrm{d})}$ and a
quadrupole part $\phi^{(\mathrm{q})}$,
    \begin{align}\label{phi_expansion}
    \phi_1
    =
    \phi^{(\mathrm{m})}
    +
    \phi_1^{(\mathrm{d})}
    +
    \phi_1^{(\mathrm{q})}
    \,,
    \qquad
    \phi_2
    =
    \phi^{(\mathrm{m})}
    +
    \phi_2^{(\mathrm{d})}
    +
    \phi_2^{(\mathrm{q})}
    \,,
    \end{align}
where the parts satisfy the equations:
    \begin{align}\label{eqs_for_multipoles}
    \nabla^2\phi^{(\mathrm{m})}
    &=
    -
    4\pi\delta(\vec r)
    \,,
    \nonumber\\
    \nabla^2\phi_1^{(\mathrm{d})}
    &=
    4\pi\vec r_1\cdot\nabla\delta(\vec r)
    \,,
    \hspace{2.15cm}
    \nabla^2\phi_2^{(\mathrm{d})}
    =
    4\pi\vec r_2\cdot\nabla\delta(\vec r)
    \,,
    \nonumber\\
    \nabla^2\phi_1^{(\mathrm{q})}
    &=
    -
    2\pi
    (\vec r_1\cdot\nabla)^2\delta(\vec r)
    \,,
    \hspace{15mm}
    \nabla^2\phi_2^{(\mathrm{q})}
    =
    -
    2\pi
    (\vec r_2\cdot\nabla)^2\delta(\vec r)
    \,.
    \end{align}

Representation (\ref{phi_expansion}) generates many terms in
Eq.~(\ref{eq1}). The linear term in $\vec r_2$ after substitution
into the Panofsky-Wenzel Eq.~(\ref{panofsky_wenzel}) gives rise to a
transverse impedance which corresponds to a kick on particle 2 when
both particles travel along the reference orbit without offset. We
will call this  impedance the transverse \emph{monopole} impedance
and denote the corresponding longitudinal impedance by
$Z_{\parallel,m}$, where
    \begin{align}\label{monopole_impedance}
    Z_{\parallel,m}
    &=
    \frac{1}{2\pi c}
    \int_{S_B}
    \nabla \phi_{1,B}^{(\mathrm{m})}
    \cdot
    \nabla \phi_{2,B}^{(\mathrm{d})}
    \, d S
    -
    \frac{1}{2\pi c}
    \int_{S_\mathrm{ap}}
%    \left(
    \nabla \phi_{1,A}^{(\mathrm{m})}
    \cdot
    \nabla \phi_{2,B}^{(\mathrm{d})}
%    +
%    \nabla \phi_{1,B}^{(\mathrm{m})}
%    \cdot
%    \nabla \phi_{2,A}^{(\mathrm{d})}
%    \right)
    \, d S
    \,.
    \end{align}
In many practical cases the structure geometry has both up-down and
right-left symmetry and the design orbit is on the symmetry line. In
such a case the transverse monopole impedance vanishes. The
longitudinal impedance which gives the usual transverse wake then
has a term that is proportional to the product of vector components
of $\vec r_2$ and $\vec r_1$ and one that is quadratic in $\vec r_2$
(no dependence on $\vec r_1$). According to accepted terminology, we
will call the former (in the case of impedance) the \emph{dipole}
component, $Z_{\parallel,d}$, and the latter the \emph{quadrupole}
component, $Z_{\parallel,q}$. The total impedance, the sum of the
two components, is given by
    \begin{align}\label{dipole_and_quad_impedances1}
    Z_{\parallel}
    =
    Z_{\parallel,d}
    +
    Z_{\parallel,q}
    \,,
    \end{align}
where
    \begin{align}\label{dipole_and_quad_impedances}
    Z_{\parallel,d}
    &=
    \frac{1}{2\pi c}
    \int_{S_B}
    \nabla \phi_{1,B}^{(\mathrm{d})}
    \cdot
    \nabla \phi_{2,B}^{(\mathrm{d})}
    \, d S
    -
    \frac{1}{2\pi c}
    \int_{S_\mathrm{ap}}
%    \left(
    \nabla \phi_{1,A}^{(\mathrm{d})}
    \cdot
    \nabla \phi_{2,B}^{(\mathrm{d})}
%    +
%    \nabla \phi_{1,B}^{(\mathrm{d})}
%    \cdot
%    \nabla \phi_{2,A}^{(\mathrm{d})}
%    \right)
    \, d S
    \,,
    \end{align}
and
    \begin{align}\label{quad_impedance}
    Z_{\parallel,q}
    &=
    \frac{1}{2\pi c}
    \int_{S_B}
    \nabla \phi_{1,B}^{(\mathrm{m})}
    \cdot
    \nabla \phi_{2,B}^{(\mathrm{q})}
    \, d S
    -
    \frac{1}{2\pi c}
    \int_{S_\mathrm{ap}}
%    \left(
    \nabla \phi_{1,A}^{(\mathrm{m})}
    \cdot
    \nabla \phi_{2,B}^{(\mathrm{q})}
%    +
%    \nabla \phi_{1,B}^{(\mathrm{m})}
%    \cdot
%    \nabla \phi_{2,A}^{(\mathrm{q})}
%    \right)
    \, d S
    \,.
    \end{align}
Note that for the special case of cylindrically symmetric structures
the quadrupole component is identically equal to zero.

Equations (\ref{panofsky_wenzel}), (\ref{monopole_impedance}),
(\ref{dipole_and_quad_impedances}), and (\ref{quad_impedance})
allows one to calculate all the component of the transverse
impedance for a transition of arbitrary geometry.

%*************** new section ********************************

\section{Example impedance calculations in the optical
regime}\label{sec:examples}

%*************** new section ********************************

In this section we will show how to calculate the longitudinal and
transverse impedances for several simple cases of axisymmetric
systems using results of Section \ref{sec:transverse_imp}.

\begin{figure}[!htb]
    \centering
    \includegraphics[draft=false, width=1.0\textwidth]{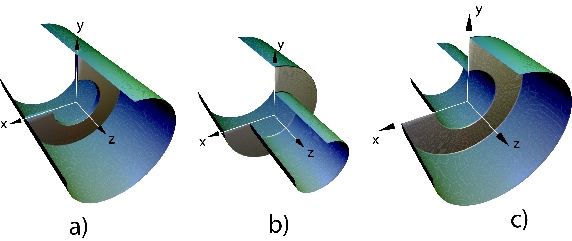}
    \caption{A short collimator (a), a step-in (b), and a step-out (c) transition
    in axisymmetric geometry. The particles move in the $+z$ direction. For the collimator $a$ denotes the
    pipe radius and $b$ the collimator radius, with $b<a$. For the step transitions,
    $a$ denotes the radius of the incoming pipe and $b$ the radius of the outgoing pipe,
    with $b<a$ for the step-in and $b>a$ for the step-out transition.}
    \label{fig:collimator_and_steps}
    \end{figure}

We first calculate the longitudinal impedance for a short, round
collimator shown in Fig.~\ref{fig:collimator_and_steps}a,
with pipe radius $a$ and collimator radius $b$ ($b<a$). In this
case both charges are located on the axis of the pipe, and the
solution to Eqs. (\ref{eqs_for_phi}), in cylindrical coordinates,
is
    \begin{align}\label{monopole_phi}
    \phi_{1,2}
    =
    -
    2
    \ln \frac{r}{a}
    \,.
    \end{align}
Substituting this solution into Eq.~(\ref{eq1}) yields
    \begin{align}\label{impedance_real4}
    Z_\parallel
    &=
    \frac{1}{2\pi c}
    \int_{S_B}
    (\nabla \phi_{1})^2
    \, d S
    -
    \frac{1}{2\pi c}
    \int_{S_\mathrm{ap}}
    (\nabla \phi_{1})^2
    \, d S
    \nonumber\\
    &=
    \frac{1}{2\pi c}
    \int_{S_B-S_\mathrm{ap}}
    (\nabla \phi_{1})^2
    \, d S
    =
    \frac{4}{c}
    \ln
    \frac{a}{b}
    \,.
    \end{align}
This is a well known result for the collimator impedance in the
high frequency limit.

The longitudinal impedance for step-in and step-out transitions
shown in Fig.~\ref{fig:collimator_and_steps}b and
\ref{fig:collimator_and_steps}c can easily be obtained from the
previous equations. In both cases the potential $\phi$ is
$-2\ln(r/a)$ and $-2\ln(r/b)$ for pipes of radius $a$ and $b$,
respectively. However, for the step-in transition, the cross-section
$S_B$ coincides with $S_\mathrm{ap}$, and the two integrals in
Eq.~(\ref{eq1a}) cancel, giving a total impedance of zero. For a
step-out transition, the difference of these integrals is the same
as given by the first line in Eq.~(\ref{impedance_real4}), and the
impedance is equal to that of a collimator with the aperture equal
to the radius of pipe B. Again, both these results are known in the
literature, and we have derived them here to demonstrate that the
optical approximation agrees with previously obtained results.

For the transverse impedance of a round collimator and a step-in
and a step-out transition we first note that only the dipole term
in Eq.~(\ref{dipole_and_quad_impedances1}) contributes to the
impedance. The quadrupole term in this equation vanishes because
the monopole potential $\phi^{(\mathrm{m})}$ in
Eq.~(\ref{quad_impedance}) has no angular dependence while the
quadrupole potential $\phi^{(\mathrm{q})}$ has an angular
dependence $\propto \cos 2\theta$ (here $\theta$ is the azimuthal
angle in cylindrical coordinates).

For a round collimator, pipe A and pipe B have the same radius $a$,
hence, $\phi_{1,A}^{(\mathrm{d})} = \phi_{1,B}^{(\mathrm{d})}$ and
$\phi_{2,A}^{(\mathrm{d})} = \phi_{2,B}^{(\mathrm{d})}$. This
reduces the integration in Eq.~(\ref{dipole_and_quad_impedances}) to
one over the col\-li\-ma\-tor area
    \begin{align}\label{collimator_impedance}
    Z_{\parallel,d}
    &=
    \frac{1}{2\pi c}
    \int_{S_B-S_\mathrm{ap}}
    \nabla \phi_{1,A}^{(\mathrm{d})}
    \cdot
    \nabla \phi_{2,A}^{(\mathrm{d})}
    \, d S
    \,.
    \end{align}
We assume now that both leading and trailing particles are offset
in the direction of the $x$ axis and define function $\psi$ such
that
    \begin{align}\label{phi_thru_psi}
    \phi_{1,A}^{(\mathrm{d})} = x_1\psi_A\,,\qquad
    \phi_{2,A}^{(\mathrm{d})} =x_2\psi_A
    \,.
    \end{align}
The function $\psi_A$ satisfies the following equation
    \begin{align}\label{dipole_potentials}
    \nabla^2\psi_A
    =
    4\pi \delta'(x)\delta(y)
    \,,
    \end{align}
where the prime denotes derivative with respect to the argument,
with the boundary condition $\psi_A = 0$ at $r=a$. The solution is
    \begin{align}\label{dipole_potential}
    \psi_A
    =
    -
    2
    x
    \left(
    \frac{1}{a^2}
    -
    \frac{1}{x^2+y^2}
    \right)
    =
    -
    2
    r\cos\theta
    \left(
    \frac{1}{a^2}
    -
    \frac{1}{r^2}
    \right)
    \,.
    \end{align}
The expressions (\ref{phi_thru_psi}) are next substituted into
Eq.~(\ref{collimator_impedance}) and integrated over the
cross-sectional area of the collimator. The calculation can be
simplified if one uses the identity
    \begin{align}\label{int_by_parts}
    \int_{S_B-S_\mathrm{ap}}
    \nabla \psi_A
    \cdot
    \nabla \psi_A
    \, d S
    =
    \int_{S_B-S_\mathrm{ap}}
    \nabla\cdot
    (\psi_A
    \nabla \psi_A
    )
    \, d S
    -
    \int_{S_B-S_\mathrm{ap}}
    \psi_A
    \nabla^2 \psi_A
    \, d S
    \,.
    \end{align}
The second integral on the right hand side vanishes because
$\nabla^2 \psi_A = 0$ in any region that does not include the
point $x=y=0$, and the first one can be cast into an integral over
the edge of the collimator $r=b$
    \begin{align}\label{integration_over_boundary}
    \int_{S_B-S_\mathrm{ap}}
    \nabla\cdot
    (
    \psi_A
    \nabla \psi_A
    )
    \, d S
    =
    -b
    \int_{0}^{2\pi}
    \psi_A
    \frac{\p \psi_A}{\p r}
    \,  d\theta
    =
    \frac{4\pi}{b^2}
    \left(
    1
    -
    \frac{b^4}{a^4}
    \right)
    \,,
    \end{align}
which gives for the longitudinal dipole impedance
    \begin{align}\label{collimator_impedance1}
    Z_{\parallel,d}
    &=
    x_1 x_2
    \frac{2}{cb^2}
    \left(
    1
    -
    \frac{b^4}{a^4}
    \right)
    \,.
    \end{align}
Using now the Panofsky-Wenzel relation Eq.~(\ref{panofsky_wenzel})
we find the $x$ component of the transverse impedance
    \begin{align}\label{collimator_impedance2}
    \frac{Z_{\perp,x}}{x_1}
    &=
    \frac{2}{\omega b^2}
    \left(
    1
    -
    \frac{b^4}{a^4}
    \right)
    \,.
    \end{align}
The right hand side of this equations defines the transverse
impedance per unit offset $x_1$---this quantity is traditionally
referred to as the transverse impedance. The result
Eq.~(\ref{collimator_impedance2}) agrees with Ref.
\cite{zagorodnov06b}.

For the step-in and step-out transitions, Eqs.
(\ref{phi_thru_psi}) and (\ref{dipole_potential}) define the
dipole potentials for pipe A; the potentials for pipe B are
obtained from these formulae by exchanging $b$ and $a$,
    \begin{align}\label{phi_thru_psiB}
    \phi_{1,B}^{(\mathrm{d})}
    &=
    x_1\psi_B\,,\qquad
    \phi_{2,B}^{(\mathrm{d})} = x_2\psi_B
    \nonumber\\
    \psi_B
    &=
    -
    2
    x
    \left(
    \frac{1}{b^2}
    -
    \frac{1}{x^2+y^2}
    \right)
    =
    -
    2
    r\cos\theta
    \left(
    \frac{1}{b^2}
    -
    \frac{1}{r^2}
    \right)
    \,.
    \end{align}
Note that the difference $\Delta\psi = \psi_A - \psi_B$
corresponds to a uniform electric field in the $x$ direction
    \begin{align}
    \Delta\psi
    =
    2
    x
    \left(
    \frac{1}{b^2}
    -
    \frac{1}{a^2}
    \right)
    \,.
    \end{align}

For the step-in transition $S_\mathrm{ap} = S_B$, and after a simple
transformation Eq.~(\ref{dipole_and_quad_impedances}) can be reduced
to
    \begin{align}\label{step_in1}
    Z_{\parallel,d}
    &=
    -
    \frac{x_1 x_2}{2\pi c}
    \int_{S_B}
    \left(
    \nabla \Delta\psi
    \cdot
    \nabla \psi_{B}
    \right)
    \, d S
    \,.
    \end{align}
Using the same integration by parts as in Eq.~(\ref{int_by_parts})
we obtain
    \begin{align}\label{step_in2}
    \int_{S_B}
    \left(
    \nabla \Delta\psi
    \cdot
    \nabla \psi_{B}
    \right)
    \, d S
    &=
    \int_{S_B}
    \nabla
    \cdot
    \left(
    \Delta\psi
    \nabla \psi_{B}
    \right)
    \, d S
    -
    \int_{S_B}
    \left(
    \Delta\psi
    \nabla^2 \psi_{B}
    \right)
    \, d S
    \nonumber\\
    &=
    b
    \int_0^{2\pi}
    \Delta\psi
    {\left.
    \frac{\pa \psi_{B}}{\pa r}
    \right|_{r=b}}
    \, d \theta
    -
    4\pi
    \int_{S_B}
    \Delta\psi
    \delta'(x)\delta(y)
    \, d S
    \nonumber\\
    &=
    0
    \,,
    \end{align}
which means that the longitudinal dipole impedance, and hence the
transverse impedance, is equal to zero for the step-in transition.
We remind the reader that the longitudinal impedance for this case
also vanishes.

For the step-out transition $S_\mathrm{ap} = S_A$. We represent
the integral over $S_\mathrm{ap}$ in
Eq.~(\ref{dipole_and_quad_impedances}) as the difference of the
integrals over $S_B$ and the integral over $S_B - S_\mathrm{ap}$.
Then the first term in this equation combined with the integrals
over $S_B$ gives zero, as follows from the calculations for the
step-in transition. Hence
    \begin{align}\label{step-out}
    Z_{\parallel,d}
    &=
    \frac{x_1 x_2}{2\pi c}
    \int_{S_B-S_\mathrm{ap}}
    \left(
    \nabla \psi_{A}
    \cdot
    \nabla \psi_{B}
    \right)
    \, d S
    \nonumber\\
    &=
    \frac{x_1 x_2}{2\pi c}
    \int_{S_B-S_\mathrm{ap}}
    \nabla
    \cdot
    \left(
    \psi_{A}
    \nabla \psi_{B}
    \right)
    \, d S
    \nonumber\\
    &=
    \frac{x_1 x_2}{2\pi c}
    b
    \int_0^{2\pi}
    \left.
    \psi_{A}
    \frac{\pa \psi_{B}}{\pa r}
    \right|_{r=b}
    \, d \theta
    \nonumber\\
    &=
    \frac{4x_1 x_2}{c}
    \left(
    \frac{1}{a^2}
    -
    \frac{1}{b^2}
    \right)
    \,.
    \end{align}
For the transverse impedance in this case we find
    \begin{align}\label{collimator_impedance3}
    \frac{Z_{\perp,x}}{x_1}
    &=
    \frac{4}{\omega a^2}
    \left(1-
    \frac{a^2}{b^2}
    \right)
    \,,
    \end{align}
which agrees with the result of Ref.~\cite{gianfelice90p}.

%*************** new section ********************************

\section{Pillbox cavity and relation between optical and diffraction
regimes}\label{sec:pillbox_cavity}

\begin{figure}[!h]
    \centering
    \includegraphics[width=0.4\textwidth]{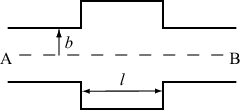}
    \caption{An axisymmetric pillbox cavity.
    \label{fig:pillbox}}
    \end{figure}

Let us consider now the pillbox cavity shown in Fig.
\ref{fig:pillbox}.
Diffraction theory gives the longitudinal impedance for a cavity
as (see, {\it e.g.} \cite{chao93})
    \begin{align}\label{diffraction_impedance}
    Z_{\parallel,\mathrm{diffraction}}
    =
    \frac{2(1+i)}{\pi^{1/2}}
    \sqrt{\frac{l}{cb^2\omega}}
    \,,
    \end{align}
where $b$ is the pipe radius and $l$ is the length of the cavity.

The optical theory predicts zero impedance for the pillbox cavity.
Indeed, in this case, all the three cross-sections $S_A$,
$S_\mathrm{ap}$ and $S_B$ are equal  (regarding the determination of
$S_\mathrm{ap}$ in this case see the footnote on page
\pageref{S_ap}), and Eq. (\ref{eq1a}) immediately gives a zero
result. The reason for the optical approximation not reproducing the
result of the diffraction theory is that Eq.
(\ref{diffraction_impedance}) corresponds to the next order
approximation in the small parameter $\lambdabar l/b^2$, which is
beyond the applicability limit of the optical regime. Indeed, if we
take the ratio of the impedance Eq. (\ref{diffraction_impedance}) to
a typical optical impedance $Z_{\parallel,\mathrm{optical}} \sim
1/c$ [see, {\it e.g.} Eq.~(\ref{impedance_real4})], we obtain
    \begin{align}
    \frac{Z_{\parallel,\mathrm{diffraction}}}{Z_{\parallel,\mathrm{optical}}}
    \sim
    \sqrt{\frac{lc}{b^2\omega}}
    \sim
    \sqrt{\frac{l\lambdabar}{b^2}}
    \,.
    \end{align}
Hence the parameter $\sqrt{{l\lambdabar}/{b^2}}$ indicates the
accuracy of the optical approximation: one can expect that
corrections to the optical regime due to diffraction effects will
be on the order of this parameter\footnote{For the geometries
considered in Section \ref{sec:examples}, the length parameter $l$
should be defined as $l\sim b$, and the accuracy of the optical
approximation is on the order of $\sqrt{{\lambdabar}/{b}}$.}.

Note that one can find in the literature references to the
impedance of the collimator given by Eq.~(\ref{impedance_real4})
as a diffraction impedance (the authors of this paper have also
used this terminology in the past). The terminology introduced in
this paper distinguishes the optical regime from the diffraction
regime, and we believe that it better describes the physics
involved.

%*************** new section ********************************

\section{Conclusion}\label{sec:summary}

%*************** new section ********************************

In this paper we have introduced an optical approximation into the
theory of impedance calculation, valid in the limit of high
frequencies. This approximation neglects diffraction effects in
the radiation process, and is conceptually equivalent to the
approximation of geometric optics in electromagnetic theory. Using
this approximation, we have derived equations for the longitudinal
impedance for arbitrary offsets of the source and test particles
with respect to a reference orbit. With the help of the
Panofsky-Wenzel theorem we have also obtained expressions for the
transverse impedance (also for arbitrary offsets). We further
simplified these expressions for the case of the small offsets
that are typical for practical applications. Our final expressions
for the impedance, in the general case, involve two dimensional
integrations over various cross-sections of the transition.

We have, in addition, demonstrated for several simple examples how
our method is applied to the calculation of impedances for simple
axisymmetric geometries that have been studied in the past.
Finally, we discussed the accuracy of the optical approximation
and its relation to the diffraction regime in the theory of
impedance.

%\bibliography{d:/my_files/bibl/accel,d:/my_files/bibl/books,d:/my_files/bibl/stupakov}

\end{document}